\documentclass[conference]{IEEEtran}
\ifCLASSINFOpdf
\fi

\usepackage{array}
\usepackage{amsmath}
\usepackage{amssymb}
\usepackage{setspace}

\usepackage{graphicx}
\usepackage{multirow}
\usepackage{cite}

\usepackage[tight,footnotesize]{subfigure}
\usepackage{makecell,multirow,diagbox}
\usepackage[justification=centering]{caption}
\usepackage{caption}
\usepackage{algorithm}
\usepackage{algorithmicx}
\usepackage{algpseudocode}
\usepackage{subfigure}
\usepackage{float}

\begin{document}

\title{An Efficient Anonymous Authentication Scheme for Internet of Vehicles}
\author{\IEEEauthorblockN{Jingwei Liu\IEEEauthorrefmark{1},
Qingqing Li\IEEEauthorrefmark{1},
Rong Sun\IEEEauthorrefmark{1},
Xiaojiang Du\IEEEauthorrefmark{2},
Mohsen Guizani\IEEEauthorrefmark{3}}
\IEEEauthorblockA{\IEEEauthorrefmark{1}State Key Lab of ISN, Xidian University, Xi'an, 710071, China.\\ Email: jwliu@mail.xidian.edu.cn, 15229259171@163.com, rsun@mail.xidian.edu.cn}
\IEEEauthorblockA{\IEEEauthorrefmark{2}Department of Computer and Information Sciences, Temple University, Philadelphia, PA 19122, USA.\\
Email: dxj@ieee.org}
\IEEEauthorblockA{\IEEEauthorrefmark{3}Department of Electrical and Computer Engineering, University of ldaho, Mosocow, ldaho, USA.\\
Email: mguizani@ieee.org}
}
\maketitle

\begin{abstract}
Internet of Vehicles (IoV) is an intelligent application of IoT in smart transportation, which can make intelligent decisions for passengers. It has drawn extensive attention to improve traffic safety and efficiency and create a more comfortable driving and riding environment. Vehicular cloud computing is a variant of mobile cloud computing, which can process local information quickly. The cooperation of the Internet and vehicular cloud can make the communication more efficient in IoV. In this paper, we mainly focus on the secure communication between vehicles and roadside units. We first propose a new certificateless short signature scheme (CLSS) and prove the unforgeability of it in random oracle model. Then, by combining CLSS and a regional management strategy we design an efficient anonymous mutual quick authentication scheme for IoV. Additionally, the quantitative performance analysis shows that the proposed scheme achieves higher efficiency in terms of interaction between vehicles and roadside units compared with other existing schemes.
\end{abstract}

\IEEEpeerreviewmaketitle

\section{Introduction}
Today, the Internet of Things (IoT) is widely used in various areas, including smart transportation, smart grid, smart health, etc. Internet of Vehicles (IoV) \cite{gerla2014internet} is one of the revolutions of IoT. It develops from Vehicular Ad hoc Networks (VANETs). VANETs cannot make intelligent decisions due to lacking the capacity of processing, analyzing, and evaluating global information collected from vehicles and infrastructures. In contrast to VANETs, IoV integrates vehicles, human, things, and networks as an intelligent unit via network technologies including deep learning, fog computing, cloud computing, etc.

\begin{figure}[tb]
 \centering
 \includegraphics[width=8.5cm]{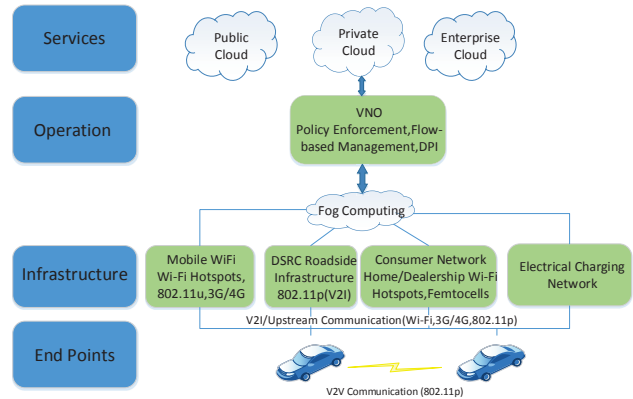}\\
 \caption{Four-level system model for IoV}
 \label{model}
\end{figure}

\begin{figure}[tb]
 \centering
 \includegraphics[width=8.5cm, height=4cm]{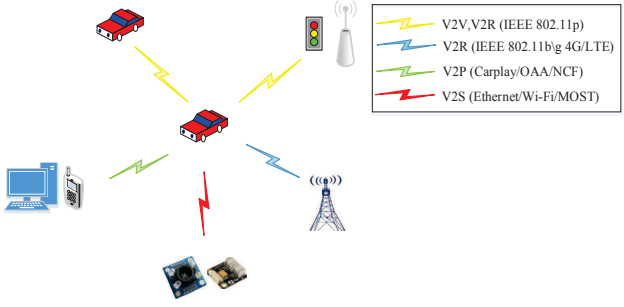}\\
 \caption{Various communication scenes in IoV}
 \label{type}
\end{figure}

Relevant scholars have proposed several reference models on IoV, such as three-level model\cite{liu2011internet}, four-level model\cite{bonomi2013smart}, and five-level model\cite{kaiwartya2016internet}. The four-level model was proposed by CISCO in 2013, as shown in Fig. \ref{model}. It mainly consists of vehicles, roadside units (RSUs), personal devices, and sensors. Various communication scenes in IoV are summarized in Fig. \ref{type}: Vehicle-to-Vehicle (V2V), Vehicle-to-Roadside unit (V2R), Vehicle-to-Personal devices (V2P) and Vehicle-to-Sensors (V2S). This kind of hybrid communication model could provide more convenient and intelligent services in IoV. The real-time connection between vehicles and IoV networks makes services more reliable and secure.

As an emerging paradigm, mobile cloud computing (MCC) is a branch of cloud computing for mobile Internet. In \cite{gerla2012vehicular}, Gerla proposed a new computing model based on MCC for vehicles---mobile vehicular cloud computing. Vehicles and RSUs often have three kinds of resources including data storage, sensors and computing. The interconnection of these resources  and Internet establishes a vehicular cloud to provide intelligent service. For instance, vehicles pick up emergency road situation, and upload it to the vehicle cloud server. Finally, the cloud server reminds the relevant vehicles to notice the breaking information. Vehicle could upload global and constant contents to the Internet. It will decreases the event processing delay. All operations are based on the cooperation of vehicular cloud, public cloud, private cloud, enterprise cloud, and big data analysis, which make IoV more intelligent.

Many previous works provide the technical basis for IoV \cite{du2005maintaining,su2017incentive,du2006secure,xiao2007survey,hui2017utility,du2007effective,huang2015cost, xiao2007internet,du2008secure,hu2005optimized,du2008security,zhang2015robust,du2009transactions}. However, IoV still faces many challenges\cite{wang2014modeling,yu2015malware,yu2015modeling,yu2015fool,manandhar2014detection}. Security threats and privacy
issues have been more and more crucial in IoV. If an attacker impersonates a vehicle to send fake messages, it may affect traveling routes of other vehicles. By now, many researches on the security of IoV \cite{liu2017vehicle,sun2017smart, rawat2017security} have been presented. Besides security concern, privacy preservation is another crucial requirement. It should prevent attackers from obtaining user's private and sensitive information, such as the user's real identity and location. However, if any vehicle is compromised, the trusted authority should be able to track it from relevant information. So, anonymity in IoV should be conditional.

Because of combing the merits of certificateless cryptosystem and short signature, certificateless short signature is suitable for recourse-constrained IoV scenario. In 2007, Huang et al. proposed the first certificateless short signature scheme and the security model \cite{huang2007certificateless}. In 2013, He et al. \cite{he2013new} proposed an efficient scheme with better  performance than the previous schemes.

In this paper, we mainly focus on the privacy preserving secure access issue in IoV. Considering the aforementioned conflicts and challenges, we propose an efficient anonymous authentication scheme for IoV. The main contributions of this paper are summarized as follows:

 \begin{itemize}
        \item The proposed scheme provides conditional anonymous mutual authentication and privacy preservation.
        \item A concept of regional management for roadside units is introduced. RSUs in the same region can work together to complete the verification of vehicles.
        \item Compared with the previous schemes, our scheme is more efficient in terms of computational overhead.
 \end{itemize}

The rest of this paper is organized as follows. In Section \uppercase\expandafter{\romannumeral2}, IoV scenario model and some preliminaries are introduced. Then, a certificateless short signature (CLSS) is proposed. In Section \uppercase\expandafter{\romannumeral4}, an anonymous authentication scheme for IoV is proposed based on CLSS. The security analysis and performance evaluation are given in Section \uppercase\expandafter{\romannumeral5}. Finally, Section \uppercase\expandafter{\romannumeral6} concludes this paper.

\section{PRELIMINARIES}
In this section, we introduce the IoV scenario model, security model, and design objectives.\par
\subsection{Scenario Model}
\begin{figure}[tb]
 \centering
 \includegraphics[width=6cm, height=6cm]{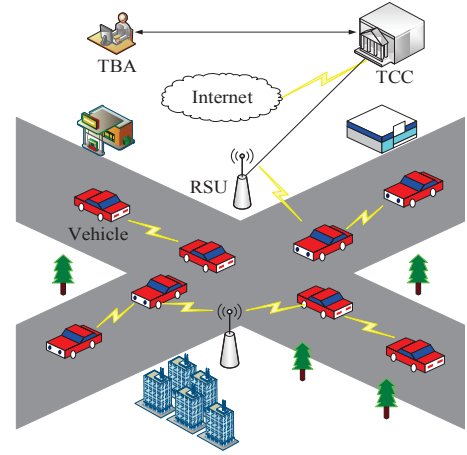}\\
 \caption{A Typical IoV Scenario}
 \label{V2R}
\end{figure}

A typical scenario model for IoV is illustrated in Fig. \ref{V2R}. It mainly consists of TCC, TBA, vehicles, and RSU.
\begin{itemize}
        \item \noindent \textbf{TCC (Transportation Control Center):} TCC is in charge of initializing systems, enrolling all entities in IoV, collecting data from RSU, tracking malicious vehicles, and maintaining revocation list.
        \item \noindent \textbf{TBA (Trace Back Authority):} TBA is responsible for receiving relevant information of dishonest vehicle, confirming malicious behavior and implementing corresponding punishment.
        \item \noindent \textbf{Vehicles:} Each vehicle in IoV is equipped with an OBU that can periodically send relevant road safety information to other vehicles and RSUs through wireless channels. In addition, it can receive and report the other OBUs' messages in a multi-hop way.
        \item \noindent \textbf{RSU (Road-Side Unit):} RSUs are the fixed road infrastructures deployed on road-side. RSUs generally communicate with TCC through wired channel. They are responsible for collecting, uploading and distributing real-time traffic information. Because RSUs can manage messages in their ranges, so they can act as gateways and provide wireless services for OBUs.
\end{itemize}

\subsection{Security Model}
In general, a CLSS contains six parts: Setup, Partial-Private-Key-Extract, Set-Secret-Value, Set-Public-Key, Signing, and Verification. We assume that there are two types of opponents to try to attack CLSS based on the ability of the master key: $A_{\uppercase\expandafter{\romannumeral1}}$ can replace any user's public key without the master key; $A_{\uppercase\expandafter{\romannumeral2}}$ can obtain the master key, but is unable to replace any user's public key. It will be proved that our scheme is existentially unforgeable against adaptive chosen message and ID attacks for two adversaries in random oracle model.

To prove the security of CLSS, we assume the following hard problems:

\noindent \textbf{Definition 1.} The k-bilinear Diffie-Hellman inversion (k-BDHI) Problem: Given two groups $G_1$ and $G_2$, and a generator P of $G_1$, a (n+1)-tuple $(P,\alpha P, \alpha^2 P,..., \alpha^n P) \in G_{1}^{n+1}$, compute $e(P, P)^{\alpha ^{-1}}$.

\noindent \textbf{Definition 2.} The k-Collision Attack Algorithm (k-CAA) Problem: Given a fixed and known integer $k$, and a (2k+2)-tuple $(t_{1},t_{2},...,t_{k},P,Q=sP,\frac{1}{t_{1}+s}P,...,\frac{1}{t_{k}+s}P) \in Z_{q}^{k}\times G_{1}^{k+2}$, output a pair (A,c) such that $A=\frac{1}{c+s}P$.

\subsection{Design Objectives}
The design objectives of our scheme are described as follows:
 \begin{itemize}
        \item \noindent \textbf{Anonymous authentication:} Anonymous authentication is an efficient approach to protect vehicle's privacy. The proposed scheme should be able to verify if the traffic information is released by legitimate vehicles. Furthermore, it should prevent attackers from obtaining vehicle's actual identity.
        \item \noindent \textbf{Conditional Privacy Preservation:} If vehicles follow the scheme honestly, their privacy should be protected very well. On the contrary, if dishonest vehicles deliberately release fake messages, TCC should be able to disclose their real identities.
        \item \noindent \textbf{Non-reputation:} OBUs and RSUs cannot deny that they have distributed the relevant traffic information.

 \end{itemize}

\section{An improved Certificateless Short Signature}
In this section, we propose an improved CLSS as an essential cryptographic primitive for our anonymous authentication in IoV.

\subsection{System Setup}
Key generation centre (KGC) first initializes the whole system as follows:

 \begin{itemize}
        \item [-] KGC takes the security parameter $l$ as input, and outputs cyclic additive group $G_{1}$ and multiplicative group $G_{2}$ with same order q, and a bilinear map $e:G_{1}\times G_{1}\rightarrow G_{2}$;
        \item [-] KGC selects a generator $P\in G_{1}$, a random system master key $s\in Z_q^*$, and computes the system public key $P_{pub} = \left\{P_{pub1}, P_{pub2} \right\} = \left\{sP, g^s\right\}$, where $g = e(P, P)$;
        \item [-] KGC chooses three one-way hash functions $H_0:\{0,1\}^*\times G_2\rightarrow Z_q^*$, $H_1:\{0,1\}^*\times G_1\rightarrow Z_q^*$, $H_2:\{0,1\}^*\times\{0,1\}^*\times G_1\times G_1\rightarrow Z_q^*$.
 \end{itemize}

Next, KGC publishes $params = \{G_1, G_2, H_0, H_1, H_2, P,$ $P_{pub}, q, g, e\}$ as the system parameters and keeps system master key $s$ in secret.
\subsection{Set-Secret-Value}
A signer chooses a random number $x\in Z_q^*$ as its secret value, and computes $PK_{ID1} = g^{x^{-1}}$ as its partial public key.
\subsection{Partial-Private-Key-Extract}
KGC chooses a random $r_{ID}\in Z_q^*$, and computes $R_{ID} = r_{ID}P$, $s_{ID} = (h_0r_{ID} - h_1s)$ mod $q$, $D_{ID} = (R_{ID}, s_{ID})$, where $h_0 = H_0(ID, PK_{ID1})$, $h_1 = H_1(ID, R_{ID})$. Then, it sends $D_{ID}$ to the signer via a secure channel.

The signer can verify $D_{ID}$ via the following equation:
 \begin{equation*}
   H_1(ID, R_{ID})P_{pub1} = H_0(ID, PK_{ID1})R_{ID}-s_{ID}P
\end{equation*}
\subsection{Set-Private-Key}
The signer sets $SK_{ID} = (x, D_{ID})$ as its private key.
\subsection{Set-Public-Key}
The signer sets $PK_{ID} = \{PK_{ID1}, PK_{ID2}, PK_{ID3}\} = \{g^{x^{-1}}, R_{ID}, x^{-1}P\}$ as its public key.
\subsection{Signing}
The signer signs a message $m \in\{0,1\}^*$ using its private key $SK_{ID}$ as follows:
 \begin{itemize}
        \item [-] Compute $h_2 = H_2(m, ID, PK_{ID2}, PK_{ID3})$;
        \item [-] Compute $\sigma = (xs_{ID} + h_0h_2)^{-1}(P + xP_{pub1})$ as the signature on $m$;
        \item [-] Send $\sigma$ to the verifier.
 \end{itemize}

\subsection{Verification}
On receiving the signer's identity $ID$, public key $PK_{ID}$, message $m$, and the corresponding signature $\sigma$, the verifier does the following steps:
 \begin{itemize}
        \item [-] Compute $h_0 = H_0(ID, PK_{ID1})$, $h_1 = H_1(ID, R_{ID})$;
        \item [-] Compute $h_2 = H_2(m, ID, PK_{ID2}, PK_{ID3})$;
        \item [-] Verify the equation $P_{pub2}\cdot PK_{ID1} = e(\sigma, h_0PK_{ID2} - h_1P_{pub1} + h_0h_2PK_{ID3})$. If it holds, the signer is authenticated; otherwise, the verification fails.
 \end{itemize}

The correctness of the scheme is proved as follows:
\[\begin{array}{l}
\begin{array}{*{20}{c}}
{}\\
{}
\end{array}e\left( {\sigma ,{h_0}P{K_{ID2}} - {h_1}{P_{pub1}} + {h_0}{h_2}P{K_{ID3}}} \right)\\
 = e({(x{s_{ID}} + {h_0}{h_2})^{ - 1}}(P + x{P_{pub1}}),{h_0}P{K_{ID2}} - {h_1}{P_{pub1}}\\
\begin{array}{*{20}{c}}
{}\\
{}
\end{array} + {h_0}{h_2}P{K_{ID3}})\\
 = e({({h_0}{r_{ID}} - {h_1}s + {x^{ - 1}}{h_0}{h_2})^{ - 1}}({x^{ - 1}} + s)P, ({h_0}{r_{ID}} - {h_1}\\
\begin{array}{*{20}{c}}
{}\\
{}
\end{array}s + {h_0}{h_2}{x^{ - 1}})P)\\
 = {g^{s + {x^{ - 1}}}}\\
 = {P_{pub2}} \cdot P{K_{ID1}}
\end{array}\]
\subsection{Security Analysis of CLSS}
The proposed CLSS scheme is secure under adaptively chosen-message and ID attacks in random oracle model. The security of the CLSS scheme relies on k-BDHI and k-CAA. The security proof of our scheme is similar to the scheme \cite{tso2007efficient,cui2007efficient}. Due to the page limitation, we omit the full proof and will give the detailed security analysis in the future work.

\section{Anonymous authentication scheme for IoV}
Based on the proposed CLSS, we design an anonymous authentication scheme for IoV.

\subsection{System Initialization}
Given the security parameter $l$, TCC generates the system public key $P_{pub} = \{P_{pub1}, P_{pub2}\} = \{sP, g^s\}$ and private key $s$ according to the method in Section \uppercase\expandafter{\romannumeral3}. Then, it chooses five one-way hash functions $H_0:\{0,1\}^*\times G_2\rightarrow Z_q^*$, $H_1:\{0,1\}^*\times G_1\rightarrow Z_q^*$, $H_2:\{0,1\}^*\times\{0,1\}^*\times G_1\times G_1\times G_1\rightarrow Z_q^*$, $H_3:\{0,1\}^*\rightarrow Z_q^*$, $H_4:\{0,1\}^*\times \{0,1\}^*\rightarrow Z_q^*$. Next, TCC chooses an encryption algorithm based on elliptic curve cryptography (ECC) and a message authentication code function $MAC _{key(\cdot)}$. At the same time, TCC maintains and updates two lists: one is the legitimate user list $Ls_{lu}$, and the other is revocation list of illegal users $Ls_{rb}$.

TCC publishes $params = \{G_1, G_2, H_0, H_1,H_2,H_3, H_4,$ $ P, P_{pub}, q, g, e\}$ as the system parameters.

\subsection{Registration}
RSUs and OBUs submit their registration requests to TCC respectively. Each OBU or RSU in IoV has its own identity $ID\in \{0,1\}^*$ that is unique and is stored into the tamper-proof device of it.

\subsubsection{OBU Registration}
An OBU chooses a random $x_o\in Z_q^*$, computes its partial public key $PK_{ID1o} = g^{x_o^{-1}}$, and then sends the registration request message $req=(ID_o, PK_{ID1o})$ to TCC. Then, TCC chooses a random $r_{IDo}\in Z_q^*$, and computes $R_{IDo} = r_{IDo}\cdot P$, $s_{IDo} = (h_0r_{IDo} - h_1s)$ mod $q$, $D_{IDo} = \{R_{IDo}, s_{IDo}\}$, in which $h_0 = H_0(ID_o, PK_{ID1o})$, $h_1 = H_1(ID_o, R_{IDo})$. Finally, TCC sends $D_{IDo}$ to the OBU via a secure channel.

The OBU sets $SK_{IDo} = (D_{IDo}, x_o)$ as its private key and keeps it in secret. Then, it uses its private key to compute partial public key $PK_{ID3o} = x_o^{-1}P$, and sets $PK_{IDo} = \{PK_{ID1o}, PK_{ID2o}, PK_{ID3o}\} = \{g^{x_o^{-1}}, R_{IDo}, x_o^{-1}P\}$ as its public key.

Verification on $D_{IDo}$: OBU verifies $D_{IDo}$ by $H_1(ID_o,R_{IDo})P_{pub1} = H_0(ID_o, PK_{ID1o})R_{IDo} - s_{IDo}P$. If the equation holds, the OBU accepts $D_{IDo}$ as its partial private key; otherwise, the OBU rejects the partial private key and aborts.

After completing the registration of the OBU, TCC adds relevant information to $Ls_{lu}$.

\subsubsection{RSU Registration}
On receiving the registration request $req = ID_R$ from a RSU, TCC computes $h_3 = H_3(ID_R)$, $D_{IDR} = \frac{1}{h_3 + s}P$, and sends $D_{IDR}$ to RSU via a secure channel.

Verification on $D_{IDR}$: RSU verifies $D_{IDR}$ by $e(D_{IDR}, H_3(ID_R)P + P_{pub1}) = g$. If the equation holds, RSU accepts $D_{IDR}$. Then RSU applies to TCC for revocation list $Ls_{rb}$.

\subsection{Report Uploading}
This process can be divided into two phases: pseudonym generation and report signing.

\subsubsection{Pseudonym Generation}
In this part, we introduce a concept of regional management for RSUs. RSUs in the same area are equipped with the same public/private key pairs. TCC periodically generates public/private key pairs, and issues them to RSUs within its range via a wireless secure channel.

When a vehicle enters a new area, it will receive the broadcasted public key from a RSU. If the vehicle wants to enjoy the service provided by this RSU, it needs to send a access report to the RSU. Then, it utilizes the public key to generate a pseudonym $f = E_{pk}(r\|ID_o)$ from user's each report $r$.

\subsubsection{Report Signing}
OBU performs the following steps to complete report signing:
 \begin{itemize}
        \item [-] Obtain a current time stamp  $T\in \{0,1\}^*$;
        \item [-] Choose a random $t\in \{0,1\}^*$, compute $PK_{IDo}' = \{PK_{ID1o}', PK_{ID2o}', PK_{ID3o}'\} = \{th_0PK_{ID1o}, $ $tPK_{ID2o}, tPK_{ID3o}\}$, $P_{pub1}' = tP_{pub1}$, and broadcast $PK_{ID1o}'$ and $P_{pub1}'$ to the other entities within its range;
        \item [-] Compute the following equations:
         \begin{eqnarray}
           h_0 &=& H_0(ID_o, PK_{ID1o}) \\
           h_1 &=& H_1(ID_o, R_{IDo}) \\
           h_2 &=& H_2(r, T, f, PK_{ID_{2o}}', PK_{ID_{3o}}')\\
           r1  &=& (t\cdot h_0)\oplus h_2, r2 = (t\cdot h_1)\oplus h_2
         \end{eqnarray}
        \item [-] The signature on report $r$ is calculated as follows:
        \begin{eqnarray}
          \sigma &=& t^{-2}(x_os_{IDo} + h_0h_2)^{-1}(P + x_oP_{pub1})
        \end{eqnarray}
        \item [-] Send the service request message $Req = (T, f, \sigma, r, r_1, r_2)$ to RSU.
 \end{itemize}

 \subsection{Mutual Authentication}
 The RSU can verify the OBU's identity $ID_o$ and report $r$. Similarly, OBU uses the message authentication code function with the shared session key to authenticate RSU.
 \subsubsection{RSU Verifies OBU}
 On receiving the service request $Req$ from OBU, RSU first checks the validity of the time stamp $T$. Then, it authenticates OBU as follows:
 \begin{itemize}
        \item [-] Compute $h_2 = H_2(r, T, f, PK_{ID_{2o}}', PK_{ID_{3o}}')$;
        \item [-] Compute $h_0' = r_1\oplus h_2$, $h_1' = r_2\oplus h_2$;
        \item [-] Verify the signature via the equation $P_{pub2}\cdot PK_{ID1o}' =e (\sigma, h_0'PK_{ID2o}' - h_1'P_{pub1}' + h_0'h_2PK_{ID3o}')\cdot h_{0}'$.
 \end{itemize}

\subsubsection{OBU Verifies RSU}
In contrast, OBU also needs to authenticate RSU.
 \begin{itemize}
        \item [-] RSU uses its private key to decrypt OBU's pseudonym: $(r\|ID_o) = D_{sk}(f)$. Next, it extracts OBU's identity $ID_o$. Then, RSU retrieves $ID_o$ in $Ls_{rb}$. If $Ls_{rb}$ contains $ID_o$, the authentication and service are terminated. After obtaining OBU's real identity, RSU computes
            $h_{ID_o} = H_3(ID_o)$, $key = H_4(h_{IDo}, ID_R)$, $mac = MAC_{key}$ $(h_{IDo})$, and sends the message authentication code $mac$ to OBU;
        \item [-] Upon receiving $mac$ from RSU, OBU computes $h_{ID_o}' = H_3(ID_o)$, $key' = H_4(h_{IDo}, ID_R)$, $mac' = $ $MAC_{key'}(h_{ID_o}')$, and checks if $mac'$ is equal to the received $mac$. If both values are equivalent, RSU is authenticated.
 \end{itemize}

\subsection{Vehicle Tracking}
If a vehicle broadcasts the false message, the prosecutor will send the vehicle's service request message $Req$ to TBA. TBA first confirms whether the vehicle is malicious. If the vehicle has malicious behavior, TBA sends the request message $Req$ and relevant evidence to TCC. Then, TCC can reveal real OBU's identity as follows:

\begin{enumerate}
  \item [-] TCC finds the corresponding RSU that provides service for the dishonest vehicle according to the vehicle location information provided by TBA.

  \item [-] TCC obtains the exact time of the dispute by checking the time stamp $T$ in the service request message $Req$.

  \item [-] TCC finds the public/private key pair $(pk, sk)$ that the dispute used. Then, it computes $D_{sk}(f) = (r\|ID_o)$.

  \item [-] RSU extracts the OBU's identity $ID_o$ from $Req$.

  \item [-] TCC adds the dishonest vehicle identity $ID_o$ into the revocation list $Ls_{rb}$, and updates it.

  \item [-] TCC sends the malicious vehicle's identity to TBA via a secure channel. Then, TBA records the dishonest vehicle's behavior and implements corresponding punishment.
\end{enumerate}

\section{Security Analysis and Performance Evaluation}
\subsection{Security Analysis}
In this section, we analyze the security properties of the anonymous authentication scheme in the following respects.

\subsubsection{OBU Anonymity}
In our scheme, the OBU's real identity is converted into the pseudonym $f = E_{pk}(r\|ID_o)$ that is not managed by any third party.
The report $r$ makes each OBU's pseudonym one-time, so adversaries cannot distinguish if two different pseudonyms come from a same vehicle. Moreover, it is intractable for adversaries to reveal OBU's actual identity without RSU's private key $sk$.
Furthermore, the public key of a vehicle would be different after multiplied by a random $t$, so none of the public keys can be linked to the same vehicle.

In the proposed scheme, any third party cannot obtain OBU's real identity, so our scheme realizes the anonymity of the OBU.

\subsubsection{Non-repudiation}
OBU cannot deny the behavior of submitting some messages, because the service request message $Req$ includes OBU's pseudonym $f$. RSU can discover OBU actual identity by computing $D_{sk}(f)$ with its private key. Therefore, non-repudiation property is satisfied.

\subsubsection{The Security of Session Key}
In our scheme, the session key is a hash value that combines OBU's real identity with RSU's identity. The security of the session key depends on the security of OBU's identity. According to the aforementioned analysis, we find that OBU's real identity is secure. Therefore, the proposed scheme can guarantee that no third party can obtain the session key.

\subsubsection{Mutual Authentication}
RSU can authenticate OBU by verifying the CLSS signature of OBU. In our scheme, only OBU and RSU know OBU's real identity $ID_o$, so the session key $key = H_4(h_{IDo}, ID_R)$ is only shared between OBU and RSU. OBU can verify RSU by checking if the computed $mac'$ is equal to the received $mac$. Therefore, mutual authentication between the OBU and RSU is achieved.

\subsubsection{Resistant to Replay Attacks}
In our scheme, current time stamp $T$ ensures the freshness of reports. On receiving the OBU's service request message $Req$, RSU first checks if the time stamp $T$ is expired. If it is, RSU rejects to accept the OBU's request. Thus, our scheme can resist replay attacks.
\begin{table}[hbp]
\centering
\caption{Running Time of Basic Operations}\label{time consumption} 
\setlength{\extrarowheight}{0.1cm}
\footnotesize
\tabcolsep 0.05in
\begin{tabular}{c|c}
\hline
\quad \quad \raisebox{0.05cm}{Operations} \quad \quad  & \quad \quad  \raisebox{0.05cm}{Time(ms)} \quad \quad \\
\hline
Pairing &   11.88 \\
Map-To-Point    &   23.34\\
Multiplication  &   10.06\\
Exponentiation  &   10.09\\
\hline
\end{tabular}
\end{table}

\begin{table}[tbp]
\centering
\caption{Comparisons on Computation Overhead}\label{complexity} 
\setlength{\extrarowheight}{0.1cm}
\footnotesize
\tabcolsep 0.05in
\begin{tabular}{c|ccc}
\hline
\quad \quad  \raisebox{0.05cm}{Schemes}  \quad \quad &  \quad \quad \raisebox{0.05cm}{Sign}  \quad \quad &  \quad \quad \raisebox{0.05cm}{Verify} \quad \quad  & \quad \quad \raisebox{0.05cm}{Map-To-Point}  \quad \quad \\
\hline
HHC\cite{he2013new} & 1H+1M & 1H+2PP+2M & YES\\
HTH\cite{hung2016revocable} &2H+2M & 3H+4PP+1M & YES\\
THSW\cite{tso2012strongly} & 1H+1E & 1H+4PP & YES\\
CCL\cite{chen2013strong} & 1M & 1H+2PP+2M & YES\\
Our Scheme & 2M & 1PP+3M  & NO\\
\hline
\end{tabular}
\end{table}

\begin{figure}[tb]
 \centering
 \includegraphics[width=7.2cm]{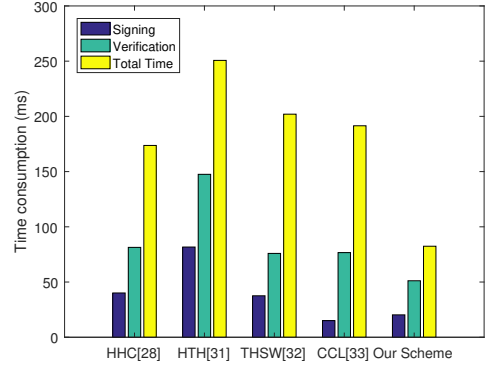}\\
 \caption{Time consumption}
 \label{consumption}
\end{figure}

\begin{figure}[tb]
 \centering
 \includegraphics[width=7.2cm]{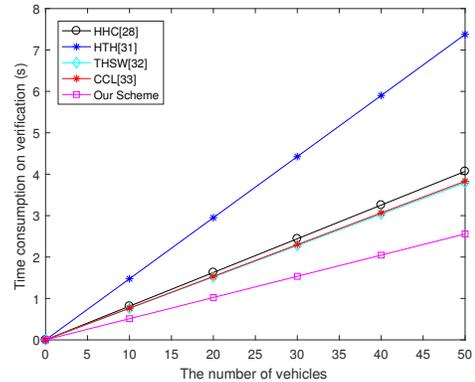}\\
 \caption{Time consumption on verification vs. the number of vehicles}
 \label{verification}
\end{figure}

\subsection{Performance Evaluation}
Due to lack of completely similar schemes for comparing, we briefly test the essential cryptographic operations instead of the whole scheme, which will not distort the results if performance evaluation. We compare our scheme with four existing schemes \cite{tso2012strongly,hung2016revocable,he2013new,chen2013strong} via experimental simulation. The simulation environment is Linux Ubuntu 16.04 LTS on an Intel Atom N450 $1.66$GHz $\times$ $ 2 $ processor. We list the running time of the basic cryptographic operations in Table \ref{time consumption}. Table \ref{complexity} shows the comparisons on computation overhead among different schemes. Let M denote multiplication in $G_{1}$, H denote the Map-To-Point operation, PP denote the bilinear pairing in $G_{1}$, and E denote the exponentiation in $G_{2}$.

In the signing phase, our scheme only requires two scalar multiplications in $G_{1}$. In the phase of verification, it requires one pairing operation and three scalar multiplications.
Fig. \ref{consumption} shows the time consumption on signing, verification and total time of these schemes. Our scheme takes the least time overhead. Fig. \ref{verification} shows the trend of the time consumption on verification with the increase of the number of vehicles. When a large number of vehicles enter the RSU's range, our scheme can provides quicker verification compared to the other schemes.

Based on test experience, in general, the energy overhead on communication is only about one-thousandth or less of that on computation, so the communication overhead is ignored in the assessment process. Therefore, as a whole, our scheme achieves better performance than the other selected schemes. It is more suitable for IoV scenarios.

\section{CONCLUSION}
In this paper, we proposed an anonymous mutual authentication scheme based on a certificateless short signature for the vehicles and RSUs in IoV. The scheme is existentially unforgeable under adaptive chosen message attack in random oracle model. The security analysis shows that the proposed mutual authentication scheme can simultaneously achieve privacy preservation and traceability of vehicles, that is conditional anonymity. Moreover, compared to the existing schemes, our scheme has lower computation overhead and achieves higher efficiency. So it is an efficient conditional anonymous authentication solution for IoV scenes.

\section*{Acknowledgment}
This work is supported by Natural Science Basic Research Plan in Shaanxi Province of China (No. 2016JM6057), the 111 Project (B08038) and Collaborative Innovation Center of Information Sensing and Understanding at Xidian University.

\bibliographystyle{IEEEtran}
\bibliography{ms}

\end{document}